# Unfolding the complexity of quasi-particle physics in disordered materials

Sai Mu[1*†], Raina Olsen[1*], B. Dutta[2*], L. Lindsay[1], G. D. Samolyuk[1], T. Berlijn[3,4], E. D. Specht[1], K. Jin[1], H. Bei[1], T. Hickel[2], B. C. Larson[1], and G. M. Stocks[1]

[1] Materials Science and Technology Division, Oak Ridge National Laboratory, Oak Ridge, TN, 37831, USA
[2] Max-Planck-Institut für Eisenforschung GmbH, D-40237 Düsseldorf, Germany
[3] Center of Nanophase and Material Science, Oak Ridge National Laboratory, Oak Ridge, TN, 37831, USA
[4] Computational Science and Engineering Division, Oak Ridge National Laboratory, Oak Ridge, TN, 37831 USA

Keywords: *3d*-transition metal alloys; force constant disorder; phonon linewidths; inelastic x-ray/neutron scattering; density functional theory

AUTHOR INFORMATION

* These authors contributed equally.

† Present address: Materials Department, University of California, Santa Barbara, California, 93106 USA.

**Abstract**

The concept of quasi-particles forms the theoretical basis of our microscopic understanding of emergent phenomena associated with quantum mechanical many-body interactions. However, quasi-particle theory in disordered materials has proven difficult, resulting in the predominance of mean-field solutions. Here we report first-principles phonon calculations and inelastic x-ray and neutron scattering measurements on equiatomic alloys (NiCo, NiFe, AgPd, and NiFeCo) with force constant dominant disorder – confronting a key 50-year-old assumption in the Hamiltonian of all mean-field quasi-particle solutions for off-diagonal disorder. Our results have revealed the presence of a large, and heretofore unrecognized, impact of local chemical environments on the distribution of the species-pair-resolved force constant disorder that can dominate phonon scattering. This discovery not only identifies a critical analysis issue that has broad implications for other elementary excitations such as magnons and skyrmions in magnetic alloys, but also provides an important tool for the design of materials with ultra-low thermal conductivity.

Quasi-particle elementary excitations including electron quasi-particles, phonons, magnons, plasmons, excitons, etc., along with skyrmions[1,2], Majorana fermions[3] and their mutual interactions, represent remarkably successful theoretical descriptions of emergent phenomena associated with quantum mechanical many-body interactions. For example, the seminal discovery of the quantized thermal Hall effect in the spin liquid candidate $RuCl_3$[4,5] can be explained by the coupling of quasi-particles (phonons and chiral Majorana edge modes) in Kitaev's non-Abelian spin liquid[6,7]. As such, quasi-particle physics has provided a microscopic understanding of the underlying science of phenomena ranging from the novel properties of quantum materials to the electronic and vibrational based physics in solids. In contrast to fully ordered crystals that can be described within the Bloch theorem, the broken translational symmetry in alloys associated with substitutional chemical disorder has long challenged the development of analogous robust quasi-particle theories of configurationally averaged observables that inherently lead to the finite quasi-particle lifetimes measured experimentally – even at zero temperature. Therefore, describing both the spectral dispersion and lifetimes of quasi-particles in disordered materials is fundamentally important for understanding the underlying science of condensed matter and the design of technological materials with targeted properties.

Historically, a landmark advance in treating compositional disorder was the coherent potential approximation (CPA)[8–10]. As an analytic "single-site" theory of the configurationally averaged observables that restores the translational invariance of a self-consistently determined effective medium, the CPA formalism was initially designed to treat on-site (site-diagonal) fluctuations associated with the different alloying elements – different on-site orbital energies for the case of electrons, different atomic masses for the case of phonons. Effective as the original CPA was, it was quickly realized that extending the single-site CPA to include fluctuations in two-site quantities – inter-site hopping integrals for electrons; pairwise force constants for phonons – is nontrivial[11–14]. This problem is particularly acute for phonons where force constant (off-diagonal) disorder is crucial and entangled with the mass (diagonal) disorder, as illustrated in Fig. 1 (a). For electron quasi-particles it was possible to account for both diagonal and off-diagonal fluctuations by reformulating CPA within the language of multiple scattering theory[15] and *ab initio* density functional theory[16–18]. For phonon quasi-particles, no similar "simple" transformation of the underlying theory has been possible.

For phonons, the formulation of itinerant-CPA (ICPA) by Ghosh *et al.*[19] provided a substantial advance in analytic mean-field theories of phonon quasi-particle physics in disordered materials based on the augmented space formalism of Mookerjee[20]. The theory was demonstrated initially for the cases of NiPd and NiPt that contain large mass disorder as well as force-constant disorder[19], and has been applied subsequently to multiple binary alloys with similarly dominant mass disorder[21–24]. Remarkably, with the exception of a limited study of face-centered cubic (fcc) $Ni_{88}Cr_{12}$ and *bcc* $Fe_{53}Cr_{47}$ using an augmented space recursion (ASR) approach[25,26], neither systematic theoretical calculations of phonon dispersions and linewidths nor corresponding experimental measurements have been performed for materials exhibiting *strong* force constant disorder with *small* mass disorder. This unsatisfactory situation regarding experimental validation of the theory is further complicated by the form of underlying Hamiltonian upon which it is based. The Hamiltonian for the ICPA and all previous mean-field theories (including the CPA) for off-diagonal as well as diagonal disorder make use of the simplifying assumption that force constant disorder between individual AA, BB, and AB type pairs can be approximated by their global average in AB binary alloys[13,27]. Similar to lattice vibrations (phonons), the magnetic excitations (e.g. magnons) of disordered alloys are also challenging to describe due to site-diagonal (local moment) and off-diagonal (exchange interactions) disorder, as depicted in Fig. 1 (b). The crucial exchange interactions between individual AA, BB, and AB type pairs are usually approximated by their *ensemble averages* as calculated from linear response theory[28]. Unaccountably, although this 50-year old assumption within the Hamiltonian is at the core of the analytic mean-field formalisms, it has so far not been questioned.

In this work we address this knowledge gap for the first time through a combined first-principles theory and experimental measurement investigation of the phonon quasiparticle physics (dispersions and linewidths) of concentrated disordered alloys, NiCo, NiFe, AgPd, and NiFeCo with strong force constant disorder but minimal mass disorder. Remarkably, we have discovered from a*b-initio* supercell phonon unfolding (SPU) simulations[29–31] and their comparison with ICPA and experimental measurements, that force constant fluctuations considering each individual AA, BB, and AB type species-pairs far surpasses that of the usual global average force constant fluctuations. Moreover, we have shown that the source of the enhanced fluctuations is the inherent random variations in local chemical environments surrounding individual AA, BB, and AB type pairs. Accordingly, we have demonstrated that the longstanding approximation of replacing individual-pair force-constant fluctuations with their global averages in the Hamiltonian of quasi-particle mean-field theories for disordered materials must be reconsidered.

## Results

**Phonon properties of NiCo and NiFe**: Phonon dispersion and linewidth measurements for NiCo and NiFe samples were performed at room temperature using high resolution inelastic x-ray scattering (IXS) and inelastic neutron scattering (INS) measurements along the [001] and [011] reciprocal lattice directions. In addition, density functional theory (DFT) was employed to calculate the force constants of NiCo, NiFe, AgPd, and NiFeCo alloys with chemical disorder modeled by the supercell method. Phonon spectral functions and corresponding linewidths for equiatomic NiCo, NiFe, AgPd and NiFeCo alloys were extracted using both the ICPA[19] and the SPU approach[31]. We note that the analytical mean-field ICPA approach employs the *symmetry-averaged* force constant matrix for a pair of atoms, i.e., one single matrix for each of $\Phi^{AA}$, $\Phi^{BB}$ and $\Phi^{AB}$ in binary A-B alloys, while the nonanalytical SPU approach uses the full force constant matrix of the supercell. The SPU maps phonon bands within the Brillouin zone (BZ) of large supercells to those of the primitive BZ. While the phonon dispersion curves for an N-atom supercell form 3N continuous (sharp) phonon dispersion bands, when unfolded into the primitive BZ they map into three *effective* acoustic bands with finite-width distributions of discrete eigenvalues. This is illustrated in Fig. 1 (c, d). More experimental and calculation details can be found in the Methods section.

Figure 2 provides a color-contour overview of the phonon dispersions and phonon linewidths as obtained by SPU and ICPA calculations, along with corresponding IXS and INS measurements for disordered solid-solution NiCo and NiFe, respectively. These plots present results as a function of the wavevector $q=[\zeta_x, \zeta_y, \zeta_z]$ in reciprocal lattice units (rlu) of $2\pi/a_0$ ($a_0$ = cubic lattice parameter) for longitudinal (LA) and transverse (TA) acoustic phonons. The color-contours in Fig. 2 give the simulated phonon spectral functions overlaid by discrete phonon dispersion measurements (black solid circles). We note general agreement between first-principles phonon dispersions (SPU and ICPA) and the corresponding measurements for both NiCo and NiFe, albeit small discrepancies occur near the $[0,0,\zeta]$ BZ boundary X-point and for $TA_1$ phonons along the $[0,\zeta,\zeta]$ direction.

More important for disordered materials, though, are the phonon linewidths which provide a direct probe of force constant disorder, when mass disorder is minimal. Figure 3 shows plots of the full width at half maximum (FWHM) linewidths, $\Delta$, of the SPU and ICPA spectral functions associated with the color contours in Fig. 2. These were obtained by fitting the spectral functions with Lorentzian profiles using negligible (0.05 meV) resolution. Similarly, the linewidths corresponding to the measurements in Fig. 2 (open circles with error bars denoting uncertainties) were obtained by fitting the (constant-$q$) IXS and INS spectral data with Lorentzian profiles convoluted with the $q$-resolution functions for the respective IXS and INS instruments. The IXS instrument resolution was determined directly by measurements of the elastic linewidth (1.6 meV) and the $q$-dependent INS resolution was determined by RESLIB using the Popovici method[32].

We observe first of all, significant phonon broadening for both NiCo and NiFe in Fig. 3, especially for q>0.7 rlu. This broadening implies that force constant disorder alone (in the absence of mass disorder) can cause significant phonon scattering and, hence, lead to shorter phonon lifetimes corresponding to substantial linewidth broadening. Furthermore, the measured linewidths are not only larger for NiFe than for NiCo, but dramatically so for $q <\sim 0.5$ rlu. The linewidths for NiCo tend to fall below ~0.5 meV for $q < 0.5$ rlu while the linewidths for NiFe range from 1.5-2 meV down to $q$ as low as ~0.2 rlu, the importance of which will be discussed below.

Focusing on the simulations, we observe in Fig. 3 (a) that (within the measurement uncertainties) both the SPU and ICPA simulations agree with the measured linewidths for NiCo, albeit ICPA tends to overestimate those of LA phonons in the [0,0,$\zeta$] direction near the BZ boundary, and that the SPU simulations tend to underestimate the linewidths in the same region. In Fig. 3 (b), it is demonstrated that the linewidth simulations for NiFe using SPU are in excellent agreement with the measured linewidths for all wavevectors sampled. Conversely, the ICPA linewidths significantly underestimate the measured linewidths for NiFe – lower by more than a factor of two for the LA phonons and lower by factors of five to ten for the TA phonons.

**Decomposition of disorder-induced phonon linewidths:** To provide physical insight into the impact of force constant disorder on phonon linewidths, we show by SPU simulations in Figs. 4(a,b) the total and pair-resolved spectral functions for selected phonon modes for NiCo and NiFe. For small $q$, [0,0,0.25] rlu, the individual-pair resolved and the total spectral functions for NiCo are quite sharp and symmetric, while for NiFe both the pair-resolved and the total spectral functions are more than twice as broad. For larger $q$, [0,0,0.75] rlu, the spectral functions for NiCo and NiFe are both broader and tend to be less symmetric, particularly for NiFe. And for $q$ close to the X-point [0,0,0.95] rlu, neither the NiCo nor the NiFe total spectral functions exhibit symmetric Lorentzian behavior. Instead, they are comprised of a sharp peak and a diffuse low-energy tail. The extended tails (together with the reduced intensity) on the low energy sides for both NiCo and NiFe are analogous to those from ICPA investigations of alloys possessing both large mass and force constant disorder[19,21–24]. The tails result from the fact that the softer like-species bonds – such as Co-Co bonds in NiCo and Ni-Ni bonds in NiFe (shown later) – do not vibrate resonantly at the high frequencies that occur for the other bonds at large wavevectors. Physically, the species-pair dependent bond strengths can be understood within the context of the local electronic structure, the strength of electron hybridization and the occupations of the bonding and antibonding states, as discussed in the supplementary note 1.3 and supplementary figure 4[33].

Comparison between the SPU calculated and the measured linewidths for NiFe and NiCo provides a direct validation of the SPU approach for predicting vibrational properties in disordered binary alloys where force constant disorder dominates. This success, now raises the question as to why the ICPA so significantly underestimates the linewidths in NiFe, given that the ICPA approach[19] was specifically designed to incorporate force constant disorder within an analytic mean-field theory.

**Importance of off-diagonal randomness:** To examine off-diagonal disorder, we have used SPU calculations to test the fundamental approximation within the present ICPA Hamiltonian for AB binary alloys[19], which includes only the inter-species disorder between the "global averaged" force constants associated with individual species-pairs ($<\Phi^{AA}>$, $<\Phi^{AB}>$, $<\Phi^{BB}>$). That is, ICPA averages over the actual distribution of force constants for individual species-pairs, *i.e.* intra-species force constant disorder. While this approximation, (in general use for mean-field quasi-particle theories) greatly simplifies the ICPA formalism and has underpinned attempts to generalize the original CPA to include off-diagonal randomness[13,27], it tacitly assumes that the local atomic environment surrounding interacting pairs does not impact the force constants for individual atomic pairs significantly. In contrast, however, we show in Figs. 5 (a,b) that the impact of local chemical environments on the species-pair force constants is significant for NiCo and dramatically so for NiFe. Only the $\Phi_{xy}$ component of the force constant matrix is shown in the figure, with the other components shown in supplementary note 1.1. Figs. 5 (a,b) give the individual (nearest neighbor) species-pair force constants for NiCo and NiFe as a function of the fractional distortion, $\delta r$, from the perfect crystal bond lengths. The individual force constants vary in two important aspects: (1) they tend to become softer for increasing bond lengths $\delta r$ according to the expected bond length proportionality, and (2) the force constants vary strongly in strength (in a statistical manner) for each bond length $\delta r$ due to random local chemical environment effects. The histograms along the left vertical axes of Figs. 5(a,b) show the statistical distributions of the $\Phi^{\alpha\beta}$ for each species-pair ($\alpha,\beta$ denote species), underscoring graphically the wider distributions of force constants for NiFe than for NiCo. Therefore, the local chemical environment clearly plays a critical role in determining species-pair force constants and their fluctuations, which in turn determine the linewidth behavior. Also shown in Figs. 5(a,b) are the average force constants $<\Phi^{\alpha\beta}>$ (denoted by the yellow symbols) that represent the force constant input to the ICPA Hamiltonian.

The thick gray SPU-A curves in Fig. 3 explicitly demonstrate the consequence of using averaged species-pair force constants. SPU-A replaces the individual AA, AB, and BB species-pair force constants of the full SPU with their averages ($<\Phi^{AA}>$, $<\Phi^{AB}>$, $<\Phi^{BB}>$), as detailed in supplementary note 1.2 [33]. Therefore, SPU-A is a supercell phonon unfolding calculation that mimics the force constant averaging of the ICPA Hamiltonian. We note that the SPU-A, the full SPU and the ICPA linewidth simulations are quite similar for NiCo. While the force-constant-averaged SPU-A calculations for NiFe depart strongly from those of SPU, they reproduce almost exactly the (anomalously low) ICPA linewidth results. These observations demonstrate conclusively that it is indeed the intra-species force constant averaging associated with the present ICPA Hamiltonian that is responsible for the breakdown in the ICPA linewidth predictions for NiFe. Conversely, the agreement between the SPU-A and the ICPA results indicates directly that the present ICPA properly accounts for the effects of off-diagonal disorder properly when it is limited to inter-species (averaged) randomness. It is important to note that the underestimated ICPA linewidths are due solely to the oversimplified Hamiltonian rather than the ICPA methodology itself.

This limitation was not apparent in previous ICPA alloy studies[19,21–24,34] because the impact of the large (100-150%) mass disorder on phonon dispersion curves tended to mask the impact of force constant disorder on phonon linewidths. In any case, the results presented above indicate that the incorporation of (individual) intra-species force constant fluctuations in the ICPA phonon Hamiltonian would lead to an accurate description of phonons in disordered materials. While outside the scope of the present study, in principle the inclusion of individual-pair force constant disorder in ICPA could be achieved by discretizing intra-species force constant distributions through the introduction of additional, appropriately-weighted, ICPA components. This is similar to the discretization of the distribution of the relevant random variables (atomic displacements, spin disorder) by introducing more CPA components for configurational averaging in the electronic structure methodology[35].

To demonstrate that the importance of retaining full force constant distributions in NiFe is not an isolated case, we performed SPU and SPU-A phonon linewidth calculations for the equiatomic 4$d$ binary AgPd and the 3$d$ ternary NiFeCo alloys, both of which have negligible mass disorder (few percent). Figure 6(a) compares the SPU-A, SPU and ICPA linewidths for AgPd, demonstrating that the linewidth behaviors of SPU-A and ICPA are again similar, both of which differ strongly from the full SPU simulations for all phonon modes. Qualitatively different from NiCo and NiFe, the species averaged force constants in AgPd are nearly equal (see Fig. 5(d)), and the intra-species force constant fluctuations vary drastically depending on the bond lengths[33]. We emphasize that the force constants in AgPd depend mainly on the bond lengths and are relatively insensitive to the chemical environment. This was demonstrated by evaluating the species-pair dependent force constants distribution for the unrelaxed structure with all atoms fixed at their ideal lattice sites, i.e. making all bond lengths identical. This resulted in narrow force constant distributions in which the magnitude of the force constants depends only on the bond length, as seen in supplementary figure 2. Moreover, even though the linewidth magnitudes predicted by the full SPU simulations are much smaller than those for either NiCo or NiFe, the fractional disagreement of the species-averaged SPU-A linewidth is similar to that for NiFe. As a more complex example we show in Fig. 6(b) SPU and SPU-A linewidth simulations for the ternary 3$d$ equiatomic alloy, NiFeCo, for which ICPA is not yet formulated. The large discrepancy between the SPU and SPU-A linewidth simulations predict that the present ICPA Hamiltonian will again break down for linewidth predictions of NiFeCo, analogous to that observed for NiFe. The SPU and SPU-A phonon spectral functions of AgPd and NiFeCo are shown in supplementary figure 3.

Analyzing further, the magnitude and the wavevector, $q$, dependence of the simulated and measured linewidths for NiCo and NiFe (see Fig. 3) provide direct quantitative insight into the force constant disorder induced phonon linewidths Δ averaged over a length scale $\ell = 2\pi/q$. In particular, linewidths of Δ = 1 meV correspond to phonon lifetimes of 0.66 ps, and $q$ = 0.25 rlu corresponds to $\ell$ ~15 Å. Accordingly, at low $q$ = 0.25 rlu, the phonon broadening is expected to be small (such as observed for LA phonons for NiCo in Fig. 3) since longer wavelength phonons are expected to be less sensitive to local force constant disorder. Surprisingly, however, for NiFe at $q$ = [0,0,0.25] rlu – corresponding to $\ell$ ~15 Å – the broadening of the LA mode is substantial and more than five times that of NiCo. This result is consistent with the observations in Fig. 5 of the remarkably broader force constant distributions in NiFe than in NiCo. Moreover, by comparing SPU linewidths for NiFe with SPU-A simulations in Fig. 3, the intra-species force constant disorder (included in SPU but not in SPU-A) plays the dominant role in the NiFe phonon broadening at low-$q$. At large-$q$ (near the X-point), where force constant disorder is resolved on the scale of the fcc cubic cell, the measured and simulated phonon broadening for both NiCo and NiFe tend to increase (see Fig. 3), albeit not necessarily monotonically.

The results of this work have far reaching implications. The discovery that local chemical environments play a dominant role in determining the impact of force constant fluctuations in disordered alloys provides both fundamental insight on the underlying science of disordered alloys and technological insight toward potential local structural engineering pathways for manipulating the microscopic vibrational physics in alloys. That is, the combination of theory and experimental measurements has demonstrated that the *ab-initio* supercell approach combined with band unfolding provides an accurate description of the vibrational physics (phonon dispersion and lifetimes) of alloys with force constant dominant disorder. And, hence, that SPU-calculated and experimentally measured linewidths can be used as predictors of phonon scattering and lattice-mediated macroscopic thermal transport in the design of novel materials such as for new thermoelectric materials.

Finally, we emphasize that the physics of quasi-particles in disordered materials revealed in the present study is applicable far beyond the scope of phonon excitations. For example, similar effects can be expected in magnetic alloys with broad, local-environment induced distributions of exchange interactions. This was demonstrated in NiFeCoCr random solid solutions[36], where the broad distributions of species-pairwise Heisenberg exchange interactions calculated for this alloy could not be adequately represented by species-pair averages. In addition, the impact of *random* (off-diagonal) interatomic exchange (*e.g.* Heisenberg exchange, Dzyaloshinskii-Moriya interaction[37,38]) can be important for magnon dispersion and lifetimes, for low energy control and magnon excitations in magnon spintronics[39], and for the helical period and dynamics of skyrmions in disordered helical magnets[40]. Accordingly, the physics of force-constant disorder revealed here within the context of phonons is transferrable not only to random alloys with both mass disorder and force constant disorder, but also to the physics of other types of quasi-particles in disordered materials in which off-diagonal disorder is dominant, *e.g.* magnons or skyrmions in magnetic alloys. Thus, the potential for dominant, local environment disorder effects such as reported for phonons in this study provides a compelling case for further investigation across the rich variety of quasi-particle physics.

**Methods**

**Experimental**:

Phonon dispersion and linewidth measurements were performed on single-phase, single-crystal, solid-solution NiCo and NiFe samples (with negligibly small mass differences of 0.4% and 4.8%, respectively) grown in an optical floating zone furnace. The measurements were performed at room temperature using high resolution inelastic x-ray scattering (IXS) and inelastic neutron scattering (INS) measurements along the [001] and [011] reciprocal lattice directions. For NiCo, IXS measurements were performed on the HERIX spectrometer at beamline 30-ID-C at the Advanced Photon Source at Argonne National Laboratory and for NiFe INS measurements were performed on the HB-3 triple-axis spectrometer at the High Flux Isotope Reactor at Oak Ridge National Laboratory (ORNL).

**Computational**:

To explore the effect of force constant disorder on the phonon properties of considered equiatomic alloys, we employed the projector augmented wave method (PAW)[41] implemented in the Vienna *ab-initio* Simulation Package (VASP)[42] to evaluate the force constants using the standard supercell technique with the help of PHONOPY software[43]. Spin polarized calculations were performed in the spin collinear state except for AgPd. Exchange and correlation were treated using the generalized gradient approximation (GGA) parameterized by Perdew, Burke, and Ernzehof[44]. The electron wave functions were expanded in a plane-wave basis set with cutoff kinetic energy set to 400 eV for all calculations. Moreover, the disordered local environment was simulated explicitly using supercells constructed from special quasi-random structure (SQS)[45] without considering short-range order. A Γ-centered 4×4×4 (3×3×3) Monkhorst-Pack k-point mesh[46] was used for Brillouin Zone (BZ) integrations in multiple 64-atom (108-atom) SQSs. The equilibrium atomic positions were obtained by optimizing the cell volume and all internal degree of freedom until the Hellmann-Feynman force on each atom is lower than 5 meV/Å. The cubic shape of the supercell is maintained while relaxing. The lattice parameters ($a_0$) of NiCo, NiFe, NiFeCo and AgPd are 3.52, 3.57, 3.55 and 4.03 Å, respectively. This is consistent with previous experiments[47]. Force constants of the supercells were calculated based on the optimized theoretical structures. The phonon spectral functions of the random alloys were calculated using the itinerant coherent potential approximation (ICPA) and supercell phonon unfolding (SPU) methods, as discussed separately later. For direct comparison with experiments, theoretical phonon spectral functions were convoluted with the *E* and *Q* resolution of the measurements. Then the same procedure that was used to extract linewidths from the experimental data was used to extract the theoretical linewidths from the convoluted theoretical phonon spectra.

**The itinerant coherent potential approximation (ICPA):** The itinerant coherent potential approximation (ICPA)[19] is a Green's function-based formalism to calculate the phonon spectra in substitutionally disordered alloys. It extends the single-site coherent potential approximation (CPA) by considering scattering from more than one site embedded in an effective medium. Hence, all relevant disorders, *i.e.*, the mass, the force constant and the environmental disorders are appropriately addressed within the ICPA. While calculating the configuration-averaged quantities, the method also ensures self-consistency, site-translational invariance and analyticity of the Green's function. The phonon frequencies and the disorder-induced linewidths were determined, respectively, from the peaks and from the widths of the coherent scattering structure factor defined as

$$S^c(\vec{q},\omega) = \sum_{s,s'} c_s c_{s'} \frac{1}{\pi} \text{Im} \langle\langle G_\lambda^{s,s'}(\vec{q},\omega^2)\rangle\rangle$$

where $\lambda$ is the normal-mode branch index, $c_s$ is the coherent scattering length for species $s$, and $\frac{1}{\pi}\text{Im}\langle\langle G_\lambda^{s,s'}(\vec{q},\omega^2)\rangle\rangle$ is the configuration averaged and thermodynamics averaged spectral function associated with the species pair $s$, $s'$.

**Supercell phonon unfolding (SPU) method:** In the supercell method for alloys, a particular finite size supercell with defects breaks the space group symmetry and leads to a shrinking BZ in reciprocal space. To recover the phonon spectra within the BZ of the primitive cell, state-of-the-art band unfolding methods have been developed for electronic problems[48,49] as well as for phonon problems[29–31]. Here we use the unfolding program developed by Ref. [31] to carry out the phonon band unfolding. According to the formula derived by Ikeda and coworkers[31], the phonon spectral function $A(\mathbf{k},\omega)$ at the wave vector $\mathbf{k}$ and frequency $\omega$ is given as: $A(\mathbf{k},\omega) = \sum_J \|\hat{p}(k)v(K,J)\|^2 \delta[\omega - \omega(K,J)]$. In this expression $\hat{p}(k)$ is the projection operator for wave vector $\mathbf{k}$ in the primitive BZ, $v(K,J)$ is the eigenvector of the dynamical matrix of the supercell for phonon mode $J$ at the $K$ point (defined in the reduced BZ), and $\omega(K,J)$ is the corresponding eigenvalue. See Ref. [31] for further details.

Finally, the SPU-A calculations were performed for a single 256-atom supercell to capture aspects of the configurational averaging inherent to ICPA, whereas each of the full SPU phonon spectral function calculations presented (except for AgPd) represent averages over several (*i.e.* six for 64 and three for 108 atom) supercells. The SPU phonon spectral function for AgPd were obtained from a single 108-atom supercell.

**Connection between ICPA and SPU:** The connection of the SPU with the ICPA can be established through the postulated ansatz that configurational averaging for an *infinite* random system can be approximated by manually averaging the observables (PDOS and spectral functions) over many *finite* SQS cells. A particular advantage of the SPU approach, however, is that it couples straightforwardly with density functional theory (DFT) computed force constants without the need for any special averaging procedure. Hence, SPU has the required capability to provide complete information on the impact of force constant disorder, i.e., the fluctuations between different atomic pairs as well as their full environment dependence. However, we note that so far, neither ICPA nor SPU has been tested for random alloys dominated by force constant disorder.

**Data availability**

The authors declare that the data supporting this study are available from the corresponding author upon request.

**Acknowledgements** This work was supported as part of the Energy Dissipation and Defect Evolution (EDDE), an Energy Frontier Research Center funded by the U. S. Department of Energy (DOE), Office of Science, Basic Energy Sciences under contract number DE-AC05-00OR22725. This research used resources of the Advanced Photon Source, a DOE Office of Science User Facility operated by Argonne National Laboratory under Contract No. DE-AC02-06CH11357. Beamline support by Ayman Said of the Advanced Photon Source and Songxue Chi of the High Flux Isotope Reactor is acknowledged. This research also used resources at the High Flux Isotope Reactor, a DOE Office of Science User Facility operated by the Oak Ridge National Laboratory. This research used resources of Oak Ridge National Laboratory's Compute and Data Environment for Sciences (CADES) and the Oak Ridge Leadership Computing Facility, which is a DOE office of Science User Facility supported under Contract DE-AC05-00OR22725. Work at MPI was supported by Deutsche Forschungsgemeinschaft (Germany) within the priority programme SPP 1599. L.L. and T.B. acknowledge support from the U. S. Department of Energy, Office of Science, Basic Energy Sciences, Materials Sciences and Engineering Division. A portion of this research, T.B., was conducted at the Center for Nanophase Materials Sciences, which is a DOE Office of Science User Facility. B.D. and S.M. acknowledge Dr. Fritz Körmann and Dr. Yuji Ikeda for fruitful discussions. S. M. is grateful to Dr. Xin Huang and Dr. Lisha Fan for graphic support.

**Author contributions:** S.M., R.O., and B.D. contributed equally to this work. R.O. performed the experimental measurements; H.B. and K.J. grew the single crystals and characterized the samples; S.M. carried out the first-principles calculations and phonon unfolding simulations; B.D. performed ICPA simulations; S.M., B.C.L., and G.M.S. wrote the paper and all authors participated in discussions and contributed materially in finalizing the paper.

**Supplementary information** accompanies the paper at the [URL]

**Competing interests:** The authors declare no competing interests.

**Correspondence:** Sai Mu (sai.mu1986321@gmail.com) and G. M. Stocks (stocksgm@ornl.gov)

45. Zunger, A., Wei, S.-H. S.-H., Ferreira, L. G. & Bernard, J. E. Special quasirandom structures. *Phys. Rev. Lett.* **65**, 353–356 (1990).
46. Monkhorst, H. J. & Pack, J. D. Special points for Brillouin-zone integrations. *Phys. Rev. B* **13**, 5188 (1976).
47. Jin, K. *et al.* Thermophysical properties of Ni-containing single-phase concentrated solid solution alloys. *Mater. Des.* **117**, 185–192 (2017).
48. Popescu, V. & Zunger, A. Effective band structure of random alloys. *Phys. Rev. Lett.* **104**, 236403 (2010).
49. Ku, W., Berlijn, T. & Lee, C.-C. Unfolding first-principles band structures. *Phys. Rev. Lett.* **104**, 216401 (2010).
# Figures:

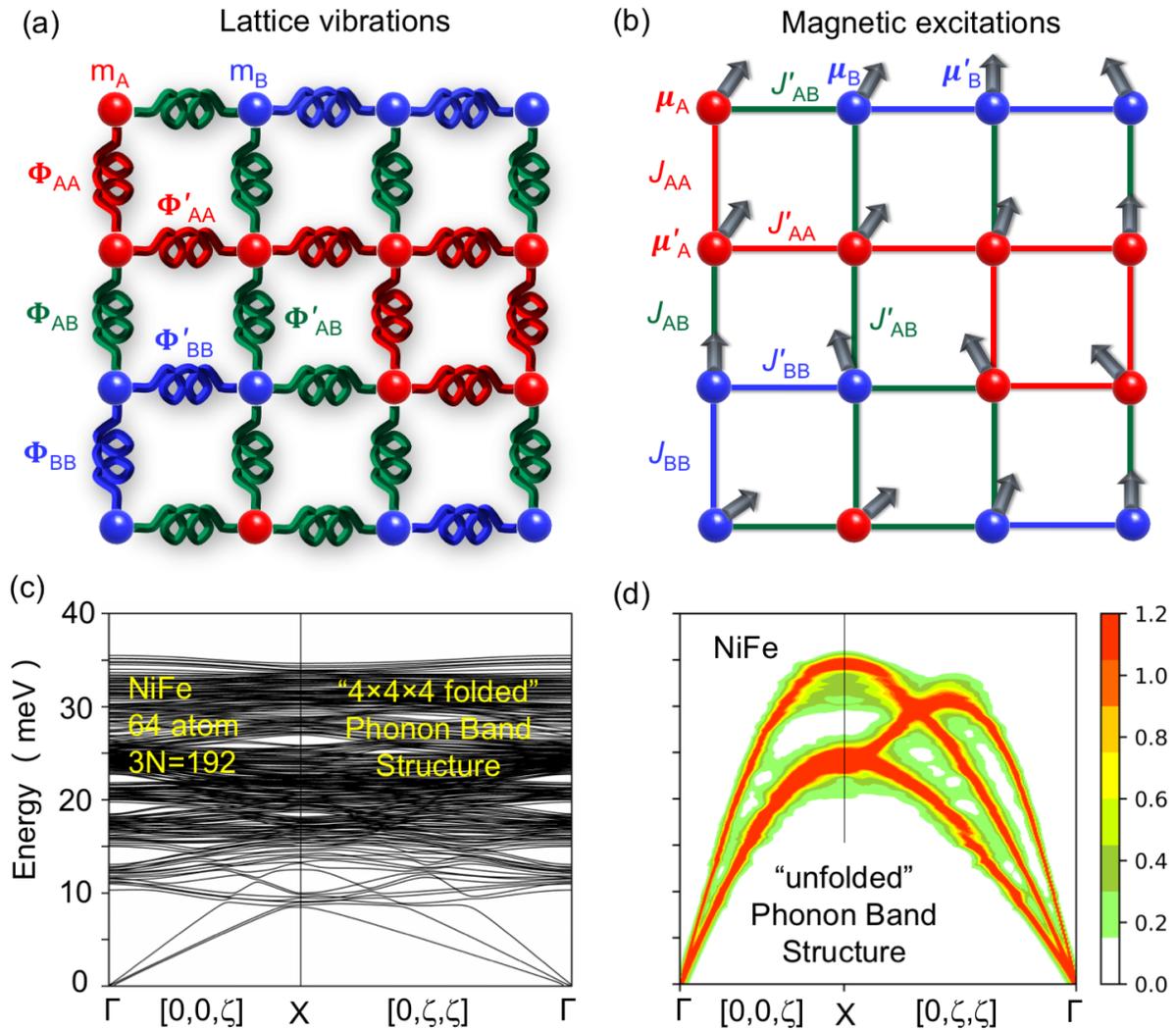

Figure 1: Site-diagonal and off-diagonal disorder for (a) lattice vibrations and (b) magnetic excitations in random solid solution alloys: $A_{0.5}B_{0.5}$. The one-site quantities are the mass ($m_i$) or local moment ($\mu_i$) on site $i$, while the two-site quantities include the force constant ($\Phi_{ij}$) or the exchange interaction ($J_{ij}$) between site $i$ and $j$. Primes indicate $\Phi'_{ij} \neq \Phi_{ij}$ due to different environments and the grey arrows in (b) denote the spin moments. (c) The folded phonon dispersion of a 64-atom 4×4×4 supercell of NiFe along [0,0,$\zeta$] and [0,$\zeta$,$\zeta$] directions, where the wavevectors are collapsed by a factor of four compared to that for a primitive cell. (d) The unfolded phonon spectral function (arbitrary units) of NiFe along [0,0,$\zeta$] and [0,$\zeta$,$\zeta$] directions.

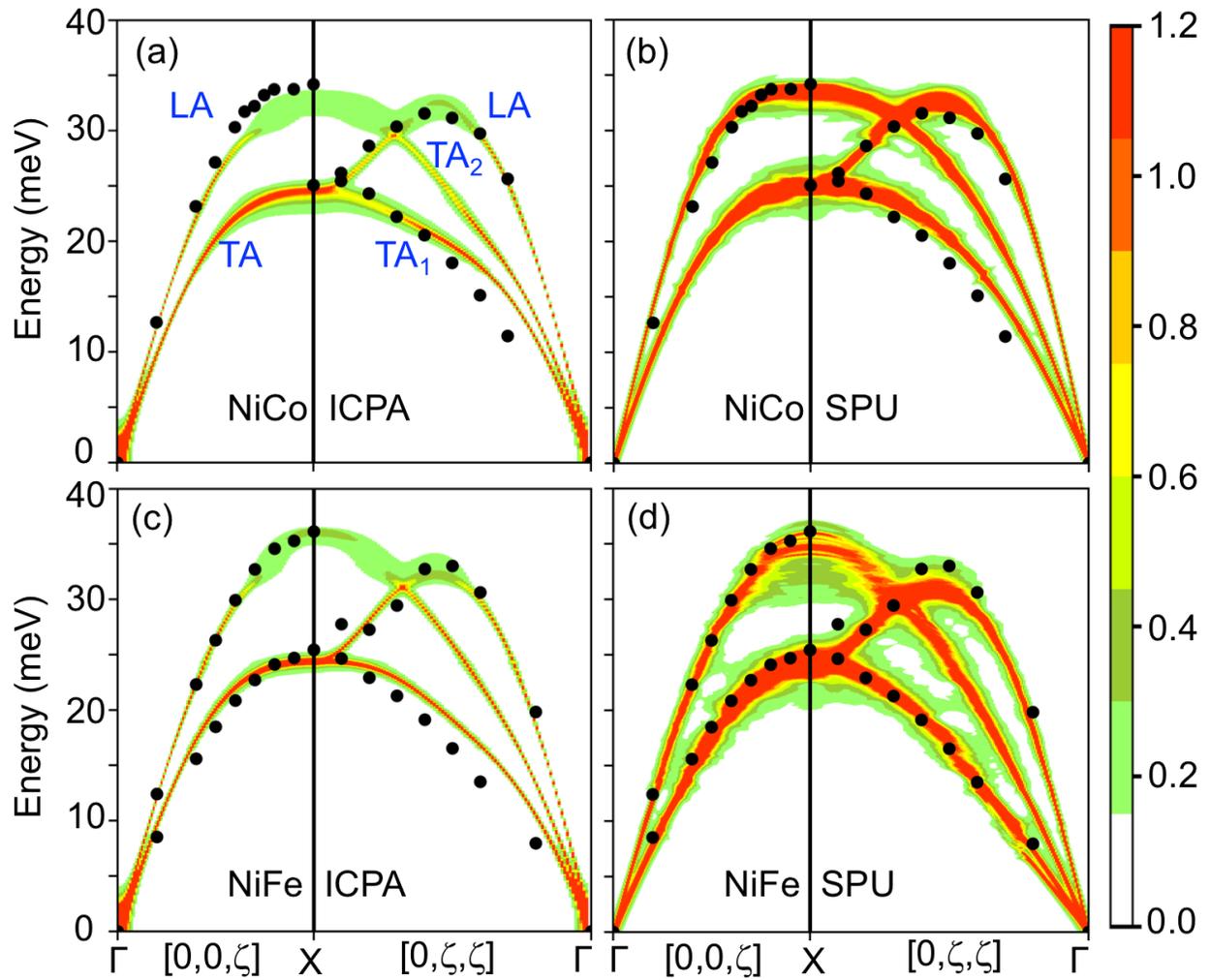

Figure 2: Phonon dispersion for NiCo and NiFe along $[0,0,\zeta]$ and $[0,\zeta,\zeta]$ directions. The measured data (solid circles) in the individual subfigures are compared with calculated phonon spectral functions (arbitrary units) from the SPU and ICPA calculations.

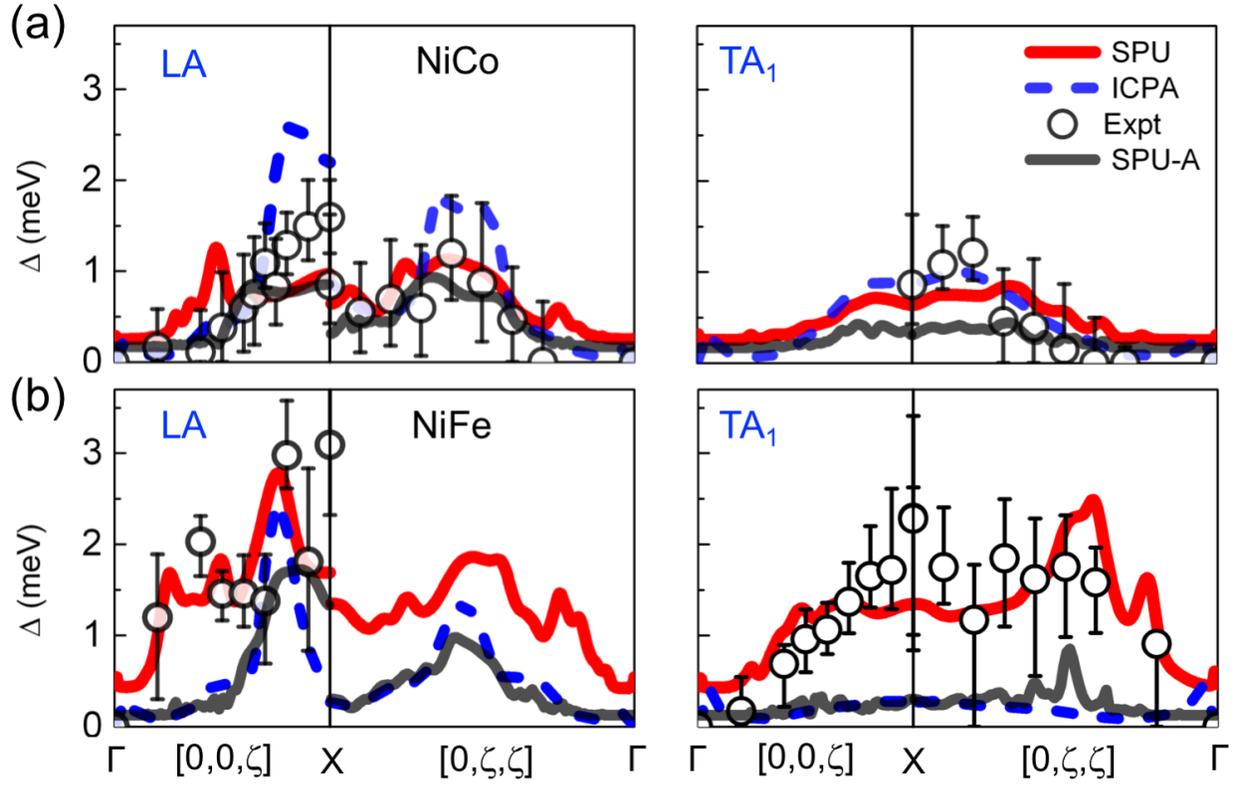

Figure 3: Phonon linewidths of NiCo and NiFe versus wavevector. Phonon broadening is quantified by the FWHM linewidths ($\Delta$) for LA and TA$_1$ modes along [0,0,$\zeta$] and [0,$\zeta$,$\zeta$] directions. The measured linewidths are denoted by the black open circles, the solid red curves are linewidths from the SPU calculations and the blue dashed curves correspond to the linewidths from ICPA. Linewidths from SPU-A are given by the translucent dark grey curves.

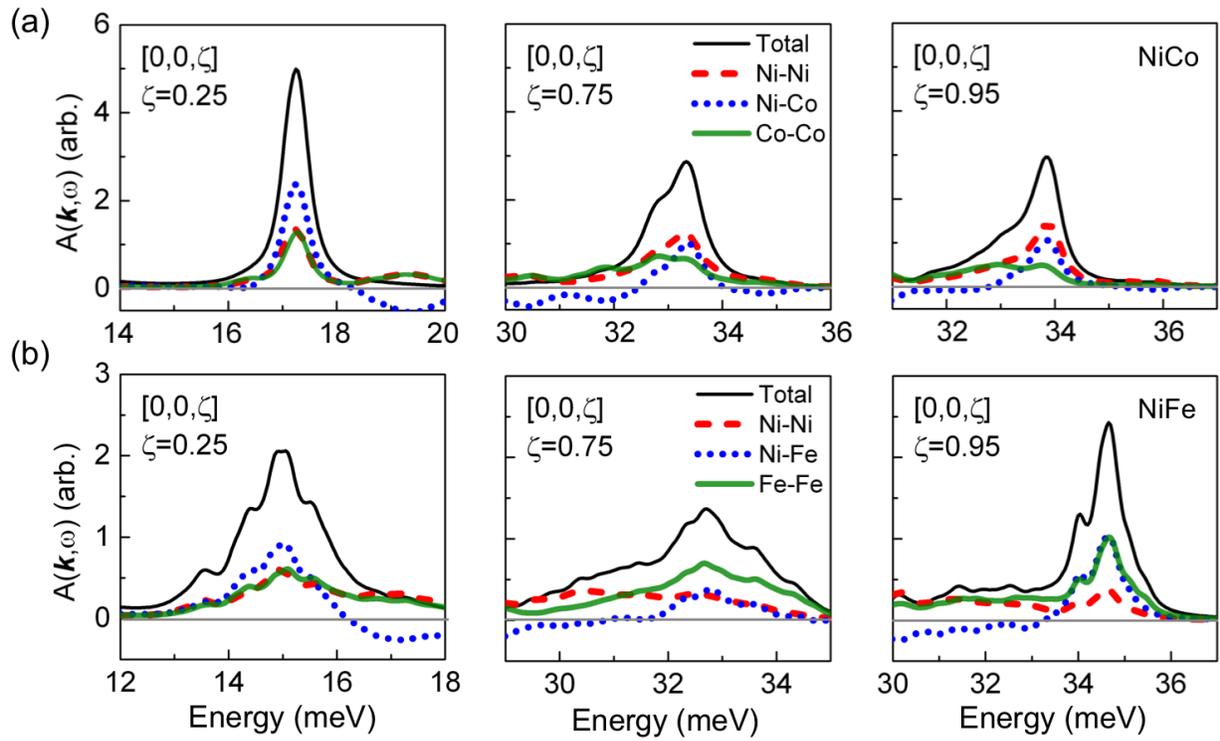

Figure 4: Total and partial spectral functions (arbitrary units) for LA modes in (a) NiCo and (b) NiFe at three $q$-points along the $[0,0,\zeta]$ direction. The grey horizontal line indicates zero.

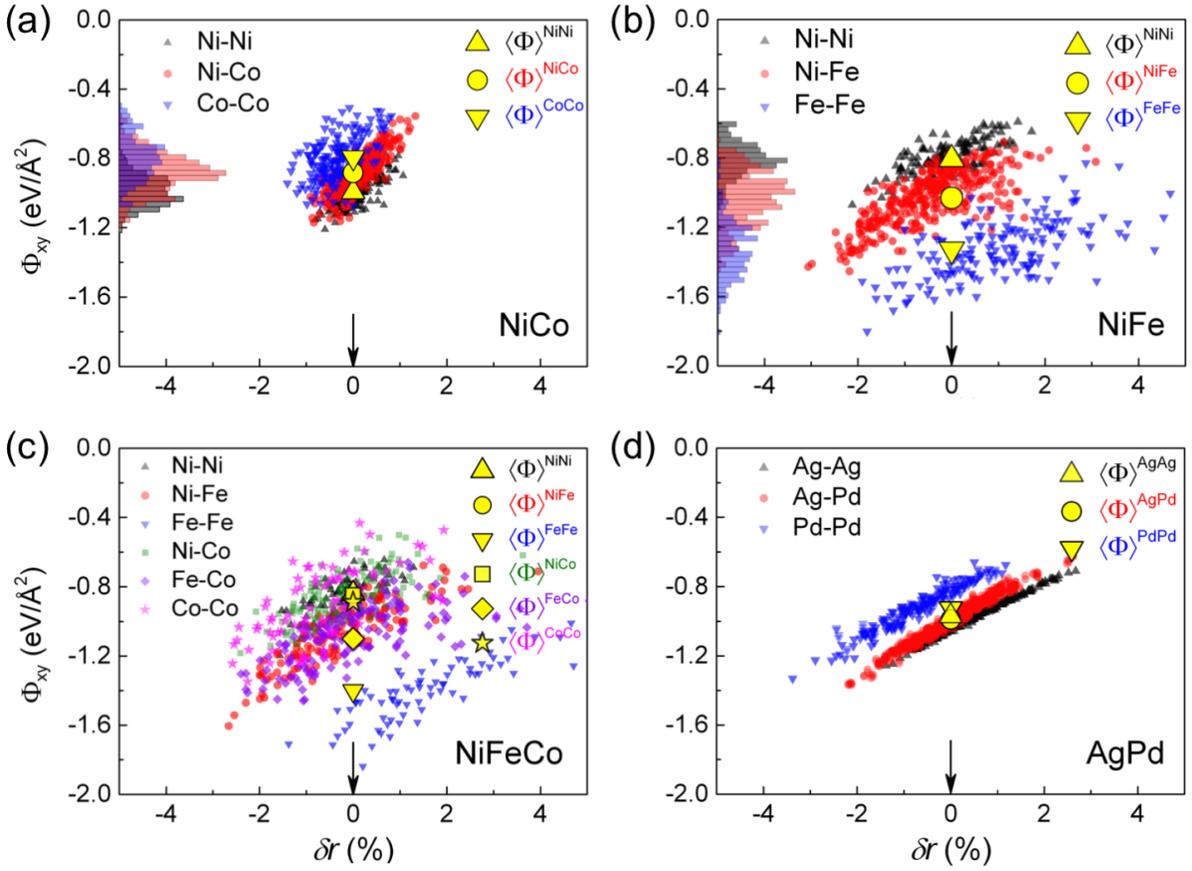

Figure 5: The *xy* component of the nearest neighbor force constant ($\Phi_{xy}$) versus percent variation of the bond lengths ($\delta r$) in (a) NiCo, (b) NiFe, (c) NiFeCo and (d) AgPd. The frequency counts of each component are shown in histograms on the y-axis. The yellow labels indicate the averaged values of the intra-species force constants.

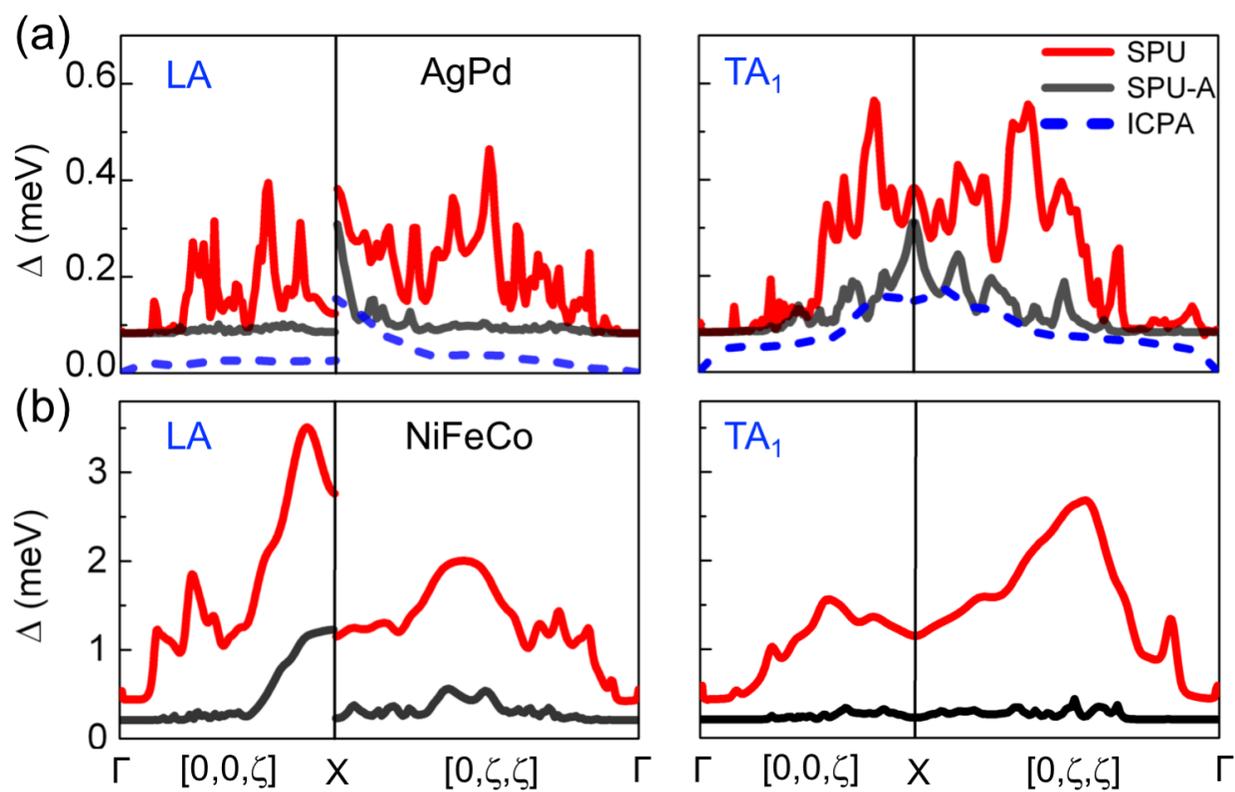

Figure 6: Calculated phonon linewidths of (a) AgPd and (b) NiFeCo.

# Supplemental Information to "Unfolding the complexity of quasi-particle physics in disordered materials"

Sai Mu[1*], Raina Olsen[1*], B. Dutta[2*], L. Lindsay[1], G. D. Samolyuk[1], T. Berlijn[3,4], E. D. Specht[1], K. Jin[1], H. Bei[1], T. Hickel[2], B. C. Larson[1], and G. M. Stocks[1]

[1] Materials Science & Technology Division, Oak Ridge National Laboratory, Oak Ridge, TN, USA
[2] Max-Planck-Institut für Eisenforschung GmbH, D-40237 Düsseldorf, Germany
[3] Center of Nanophase and Material Science, Oak Ridge National Laboratory, Oak Ridge, TN. USA
[4] Computational Science and Engineering Division, Oak Ridge National Lab, Oak Ridge, TN, USA

# Supplementary Note 1: Vibrational properties

## 1.1 Force constants versus bond lengths

The bond stiffness, or the force constants in materials are directly associated with the bond length. In the bond proportion model (BPM) (see Ref. [1] and references therein), the magnitude of the force constant increases *monotonically* and *proportionally* with reduced bond length. However, the deviation from the BPM can be understood in the following two aspects: firstly, if the bond lengths deviate strongly from the ideal interatomic spacing due to the local relaxation, the *proportionality* relationship breaks down; secondly, the force constants depend on the local chemical environment around the bonds in addition to the bond lengths. In fact, the (disordered) local environment may exert a striking effect on both the magnitude and the fluctuations of the force constants. Therefore, the monotonic relationship can be violated.

In this study, all alloying elements within a particular alloy are close to each other in the same row (either *3d* or *4d*) in the periodic table and have similar atomic size. Therefore, only small atomic displacements away from the ideal sites are observed, and the bond lengths fluctuate within 5% of the ideal interatomic spacing in all studied alloys. This is in contrast to the mixed *3d-* and *4d-* or *5d-* transition metal alloys – such as CuAu, NiPd, NiPt *etc* – whose atomic displacements are large due to large size mismatch among the alloying elements [1,2].

Figure 1 illustrates three components of the nearest neighbor force constant ($\Phi_{ij}^{\alpha\beta}$) for species-pair $\alpha\beta$ ($\alpha, \beta$=A, B within a binary AB alloy) in all studied alloys, where *i, j* are Cartesian coordinates. Figure 2 resolves each individual pairwise interaction for $\Phi_{xy}^{\alpha\beta}$. The deviations from the BPM can be viewed as due to large force constant fluctuations at the same bond length, suggesting a pronounced effect from the (disordered) local chemical environment. As for *3d*-transition metal alloys (NiCo, NiFe, NiFeCo), we found the BPM roughly holds for the NiCo alloy but starts to break down for NiFe and NiFeCo alloys, where large force constant fluctuations are observed at a certain bond lengths. If, however, all atoms are frozen on their ideal sites, we found the species-dependent force constants display similar (or even stronger) fluctuations (not shown here), indicating that the force constant is more sensitive to the chemical environment than the bond length. On the other hand, the force constants within the *4d*-transition metal AgPd alloy (with only 0.7% mass disorder) obey the BPM with little force constant variations at particular bond lengths (see Fig. 1 and 2). This suggests that the bond length plays a direct and dominant role on the force constants. Further, freezing all atoms in their ideal positions, the resulting species-resolved force constants have much weaker fluctuations, as seen in Fig. 2.

## 1.2 Comments on the averaged force constant model

Although the point group symmetry on each site of disordered alloys is broken due to random chemical environments and the resulting local atomic relaxations, we employed the symmetry operations for the ideal face-centered cubic (fcc) lattice to correlate the same type of species-dependent force constants within a particular neighboring shell. For example, all the nearest neighbor force constants are rotated so that all nearest neighbor bonds are along the (0.5,0.5,0) $a_0$ direction ($a_0$ is the lattice constant) with the help of the ideal symmetry operations, regardless of the local environments. In a perfect fcc lattice or a fully disordered solid solution, the averaged nearest neighbor force constants between any species pair for the first nearest neighbor bond (0.5, 0.5, 0) $a_0$, denoted as $\Phi$ for short, should sustain the following symmetry with only three independent components with $\Phi_{xx} = \Phi_{yy}$ and $\Phi_{xy} = \Phi_{yx}$

$$\Phi = \begin{pmatrix} \Phi_{xx} & \Phi_{xy} & 0 \\ \Phi_{xy} & \Phi_{xx} & 0 \\ 0 & 0 & \Phi_{zz} \end{pmatrix}$$

In this symmetric force constant matrix, both $\Phi_{xx}$ and $\Phi_{xy}$ are large components. Furthermore, $1/2(\Phi_{xx} + \Phi_{xy})$ gives the longitudinal force constants or central force constants (bond-stretching), while $1/2(\Phi_{xx} - \Phi_{xy})$ and $\Phi_{zz}$ gives the transverse force constants, which are much weaker. Note that this symmetry does not hold for individual force constants in alloys, but roughly holds after averaging over all the equivalent pairwise interactions, denoted as $\langle \Phi_{ij}^{\alpha\beta} \rangle$ for $i, j$ component of $\alpha\beta$ pairwise interactions $\alpha$, $\beta$ denote species in a random alloy). After averaging the force constants, if the vanishing terms – such as $\langle \Phi_{xz}^{\alpha\beta} \rangle$ and $\langle \Phi_{yz}^{\alpha\beta} \rangle$ – end up with small values, we make them zero to restore the symmetry. In addition, if $\langle \Phi_{xx}^{\alpha\beta} \rangle \neq \langle \Phi_{yy}^{\alpha\beta} \rangle$, they are replaced by $1/2(\langle \Phi_{xx}^{\alpha\beta} \rangle + \langle \Phi_{yy}^{\alpha\beta} \rangle)$. The self-interaction term is enforced assuming the acoustic sum rule (ASR). The averaged components of force constants (up to the fourth nearest neighbor shell) in NiCo, NiFe, AgPd and NiFeCo alloys are listed in Tables 1, 2, 3 and 4, respectively. The fluctuations of the nearest neighbor force constants are represented by the standard deviations of the corresponding force constant distributions (given in parentheses next to values in the tables).

Typically, the averaged force constants from one cell are not transferable to other cells. However, if the averaged force constants are from a large supercell, where statistics of the force constants are applied over many possible local environments, we assume the thermodynamic limit has been reached. The resulting averaged force constants are used as prior inputs to construct the force constants for random cells with larger sizes – sampling more configuration-dependent bonds yet losing the force constant fluctuations within each species-pair dependent (*i.e.* AA, AB, BB) force constant. This averaged force constant model based on the SPU, denoted as SPU-A, employs the same approximation made within ICPA and therefore mimics ICPA results. The unfolded phonon spectral functions for NiCo, NiFe, NiFeCo, and AgPd alloys using the full set of force constants and averaged force constants are shown in Fig. 3. SPU-A gives sharper phonon spectral functions and smaller phonon linewidths compared with those from the SPU.

### 1.3 Bond stiffness in NiFeCo, NiCo and NiFe alloys

In NiFeCo, $\langle \Phi_{FeFe} \rangle > \langle \Phi_{NiNi} \rangle > \langle \Phi_{CoCo} \rangle$. This relationship is also valid in NiFe and NiCo alloys – a binary subset of the ternary NiFeCo alloy with the same underlying fcc lattice. To understand the relative strength of the averaged species-resolved force constant, we note that the averaged intra-species force constants and the corresponding intra-species force constant fluctuations are ultimately driven by the electronic structure associated with the nature of the bonding elements and their disordered local chemical environment. More specifically, the bond strength depends on the number of electrons participating in the bonding process and the strength of the electronic hybridization – hopping integral. The former can easily be counted in the partially-filled shell, while the latter can be probed by the width of the partial electronic bands or the density of states at the Fermi level, D($E_F$).

First, the filled *3d* shell does not contribute to bonding. In NiCo, NiFe, and NiFeCo alloys whose *3d* shells in the majority-spin channel are close to fully-filled (see Fig. 4 for the density of states in NiCo and NiFe), only the electrons in minority *3d*-bands contribute to the bonding. Secondly, comparing Ni-Ni bonds with Co-Co bonds in the NiCo alloy, Ni-Ni bonds are stronger simply because there is one more *3d* valence electron from Ni in the minority-spin channel. On the other hand, for the case of NiFe, even though Ni has more *3d* electrons than Fe, the *3d* orbitals for Fe are more extended, and thereby, give larger orbital overlap. As a result, the hybridization of *3d* orbitals between Fe atoms is stronger. This is supported by the wider *3d* band width for Fe in the minority-spin channel, indicating stronger electron hybridization. Figures 4 (a-b) and (d-e) show the species-resolved partial density of states of NiCo and NiFe, respectively, calculated by the supercell method. The partial DOS clearly shows that Fe bands in the minority-spin channel are wider than Ni bands in NiFe, indicating stronger *d-d* hybridization. In addition, the band width is proportional to the inverse of the slope of the *3d* phase shifts at the band resonance ($\pi/2$), denoted as $k^{-1}$. Figures 4 (c) and (f) show *3d* phase shifts for both NiCo and NiFe alloys, respectively. The *3d* phase shift for Fe has a smaller slope ($k$(Fe)=1.06 eV$^{-1}$ compared with $k$(Ni)=2.05 eV$^{-1}$ in NiFe), and therefore Fe has a larger band width and stronger electron hybridization. Accordingly, the Fe-Fe force constant is larger than the Ni-Ni force constant in NiFe.

### 1.4 Phonon density of states (PDOS)
By projecting the eigenvectors of the phonon modes to different species, species-resolved partial

PDOS have been calculated for NiCo, NiFe, NiFeCo and AgPd alloys, as seen in Fig. 5. In the fcc lattice, there are two peaks indicating the resonance due to atomic motion; the highest frequency peak in the PDOS plot corresponds to the LA mode where the phonon band becomes flat, and the lower frequency peak relates to the top of the lower-lying TA modes. By analyzing the peak positions, the peak values, and the chemical contributions, conclusions can be drawn regarding the type of atomic motions in the energy spectrum. For example, the 8 THz resonance in NiCo is dominated by Ni atoms due to the larger intensity of the Nickel partial DOS. In addition, a 0.2 THz downward shift of the Co peak relative to the Ni peak can be seen from Fig. 5. This is due only to the relatively weaker force constants (see Table 1).

In NiFe, the relatively higher 8.2 THz resonance is dominated by the motion of Fe atoms, and the slightly lower-lying 7.8 THz resonance peak is mainly Ni-mediated. A weak shoulder remains at the 8.5 THz Fe resonance peak. The weak shoulder, corresponding to localized modes is mainly on Fe. This may smear out if the cell size is increased. Interestingly, we observe a broadening of the resonance Fe peak at about 5 THz (TA modes), forming a plateau with 1.5 THz width. The broadening of Fe-mediated peaks can be explained by the broader distribution of Fe-involving interactions (see the main text). The broadening of the resonance peaks and their mismatches from the phonon DOS provide physical insights on the phonon vibrations and the broadening of the phonon dispersion curves.

Mixing Ni, Fe, Co together, the partial DOS of NiFeCo (see Fig. 5c) sustains all the features in both NiFe and NiCo, with slight peak shifts. In the AgPd alloy, the partial DOSs of Ag and Pd ((see Fig. 5d)) are almost overlapping, indicating that Ag and Pd species in this binary alloy are almost indistinguishable.

**Supplementary Note 2: Effect of pure force disorder in NiFe**
In NiCo, the mass disorder is negligibly small (~0.4%) so the phonon broadening comes primarily from the force constant disorder. In contrast, despite the relatively small mass disorder in NiFe (~4.8%), it is still sizable compared with the mass disorder in NiCo. To isolate the pure force disorder effect in NiFe, we employ the full set of force constants of NiFe, obtained from standard *ab-initio* calculations based on a 108-atom SQS, but then calculate the phonon spectral functions by replacing all Fe masses with Co masses. The spectral functions, dispersion curves, and line widths for the actual NiFe and this "artificial NiCo" are compared in Fig. 6. As can be seen, the dispersion of the "artificial NiCo" shifts towards the low-energy region due to the slightly larger averaged mass. The phonon line widths are reduced as well due to the smaller mass disorder, but the reduction is quite weak so it has little bearing on our conclusions.

**Supplementary Note 3: Local relaxation effect on phonon properties**

Due to the small atomic size mismatch and the similar electronegativities of the alloying elements in the alloys studied here (NiCo, NiFe, NiFeCo, AgPd), the atomic displacements away from the ideal sites are small. However, NiFe displays greater local atomic displacements than those in NiCo. According to the BPM, the larger atomic displacements in NiFe accompanied by the greater bond length fluctuations, should result in larger force constant fluctuations as well. As such, to reveal the local relaxation effect on the phonon properties (dispersion curves and line widths), we employed one particular 64-atom SQS cell of NiFe and NiFeCo, and calculated the force constants with and without local relaxation, followed by unfolding to obtain the phonon properties.

The phonon spectral functions are found to be very similar regardless of whether local relaxation effects are included or not. Further detailed analysis has shown that the averaged force constants are not sensitive to the relaxation effect, as expected for these size-matched alloys. For example, the mean value and the standard deviation of the first nearest neighbor longitudinal force constants in NiFe are $\langle \Phi_\parallel^{1st} \rangle$= -0.94 eV/Å$^2$ and $\delta\Phi_\parallel^{1st}$= 0.17 eV/Å$^2$ with relaxation, compared with $\langle \Phi_\parallel^{1st} \rangle$= -0.92 eV/Å$^2$ and $\delta\Phi_\parallel^{1st}$= 0.18 eV/Å$^2$ without relaxation. The phonon line width, representing a measure of the force constant fluctuations, are found to be only slightly reduced after local relaxations. This is consistent with the fact that force constant fluctuations in 3d-transition metal alloys comes mainly from the chemical environmental configurations rather than relaxation effects. As for AgPd, the effect of local relaxation (rather than the local chemical environment) dominates force constant fluctuations as seen in Fig. 2.

## Figures

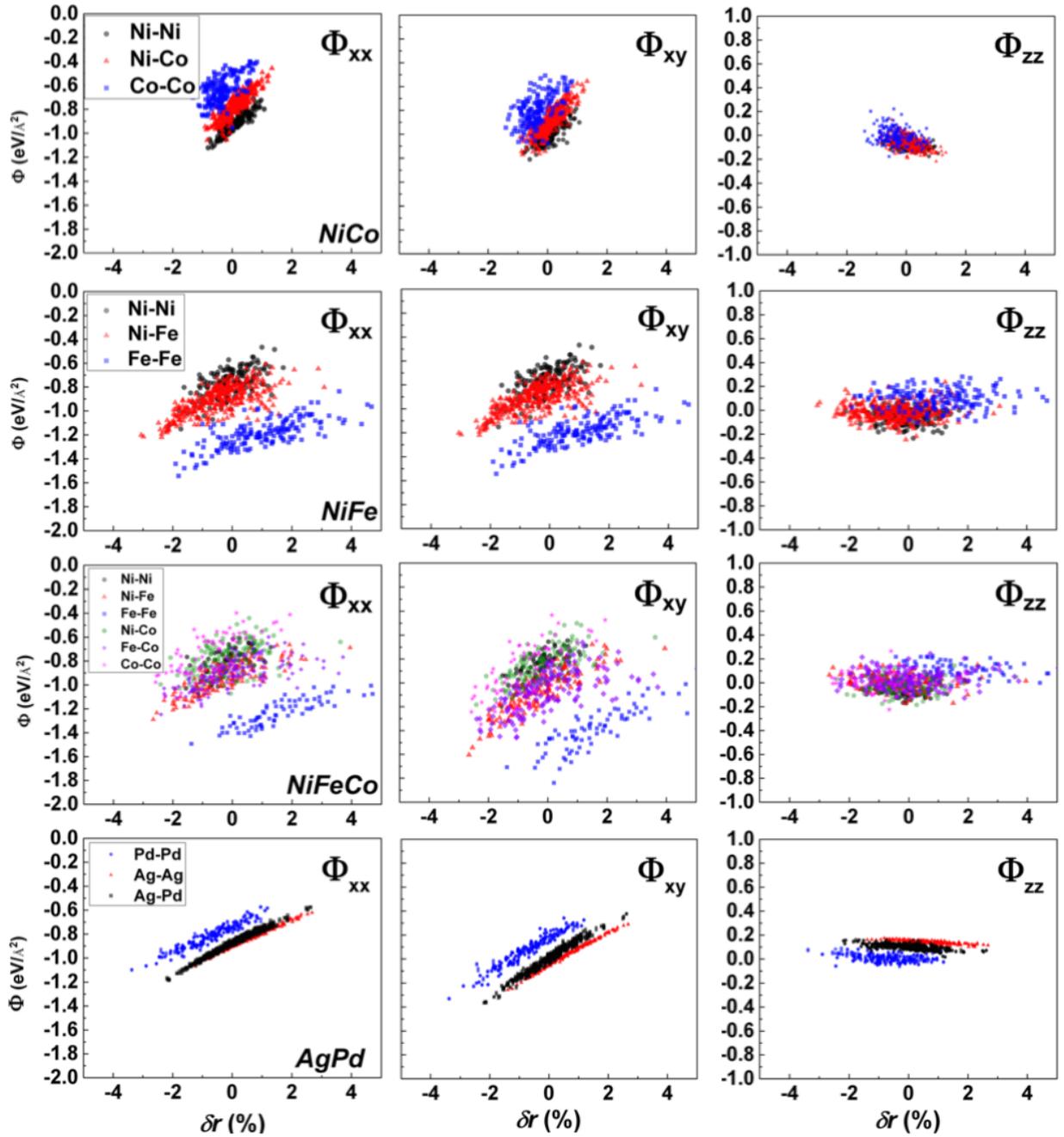

**Supplementary Figure 1** The statistics of three force constant components ($\Phi_{xx}$, $\Phi_{yy}$, $\Phi_{zz}$) within the species-resolved nearest neighbor force constants in NiCo, NiFe, NiFeCo and AgPd alloys, as a function of the percent change of ideal nearest bond lengths ($\delta r = \frac{r-r_0}{r_0}$, $r_0$ and $r$ are the ideal and actual bond lengths, respectively). All nearest neighbor bonds are rotated to align along (0.5, 0.5, 0) $a_0$, where $a_0$ is the lattice constant.

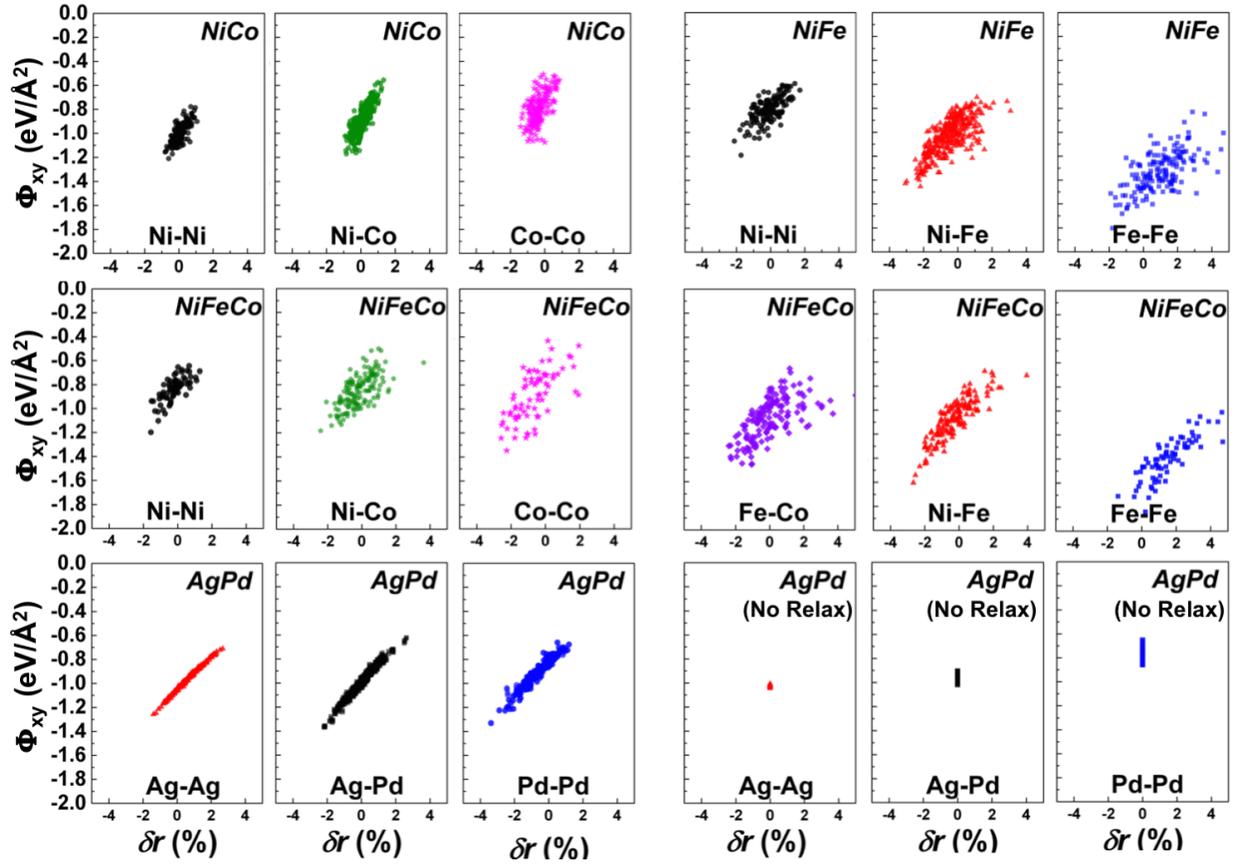

**Supplementary Figure 2** The species-resolved *xy* component of the first nearest neighbor force constants ($\Phi_{xy}^{\alpha\beta}$, where α, β denote species in a random alloy) in NiCo, NiFe, NiFeCo, and AgPd as a function of percent change of ideal nearest bond lengths ($\delta r = \frac{r - r_0}{r_0}$, where $r_0$ and $r$ are the ideal and actual bond lengths, respectively). The nearest bonds are rotated to align along (0.5, 0.5, 0) $a_0$. For AgPd, in addition to the fully relaxed force constants, the force constant distributions when atoms are clamped on their ideal sites are shown, denoted as "No Relax".

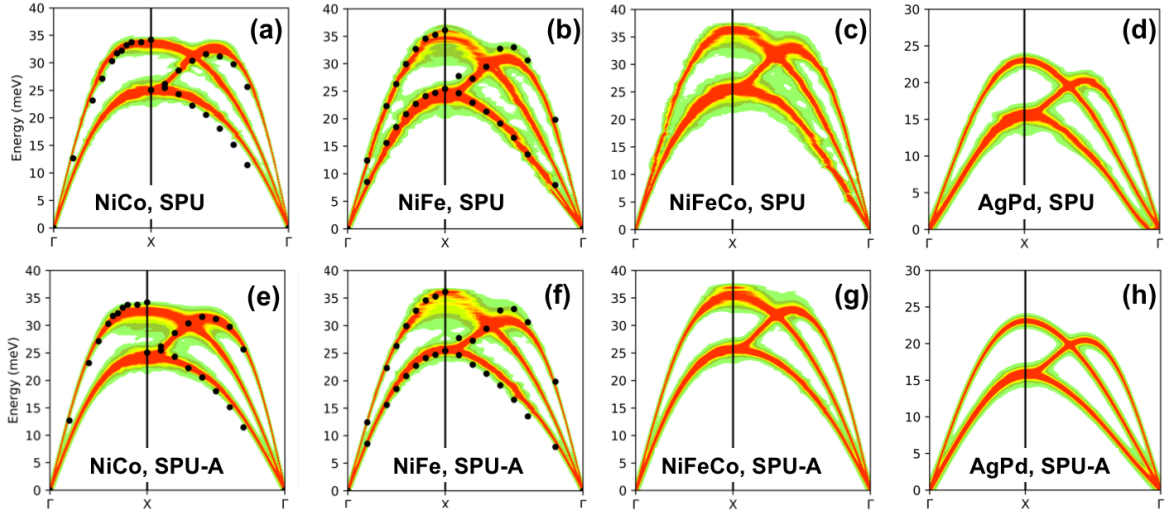

**Supplementary Figure 3** The phonon spectral functions of NiCo, NiFe, NiFeCo and AgPd using both the full set of force constants (a)-(d) that correspond to SPU results and the averaged species-resolved force constants (e)-(h) that correspond to SPU-A results (defined in the main text). Arbitrary units are employed for the phonon spectral functions. The solid black circles in the figures represent the experimental dispersion measurements.

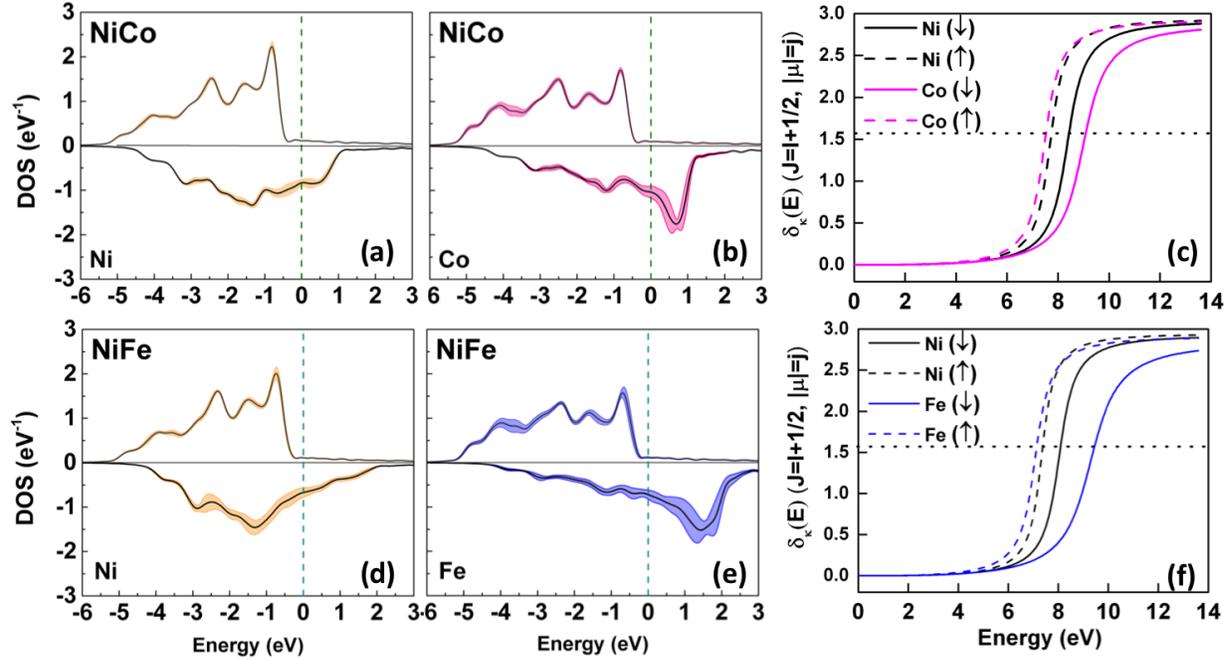

**Supplementary Figure 4** The species-resolved partial electron density of states (DOS) of NiCo (a-b) and NiFe (d-e) obtained using the supercell method. The solid black lines represent the averaged DOS in both spin channels while the colored smearing denotes the standard deviation of the DOS at each energy point. Subfigures (c) and (f) show the *3d*-phase shift (calculated from KKR-CPA) of the NiCo and NiFe alloys, respectively. Majority-spin and minority-spin channels

are denoted by ↑ and ↓.

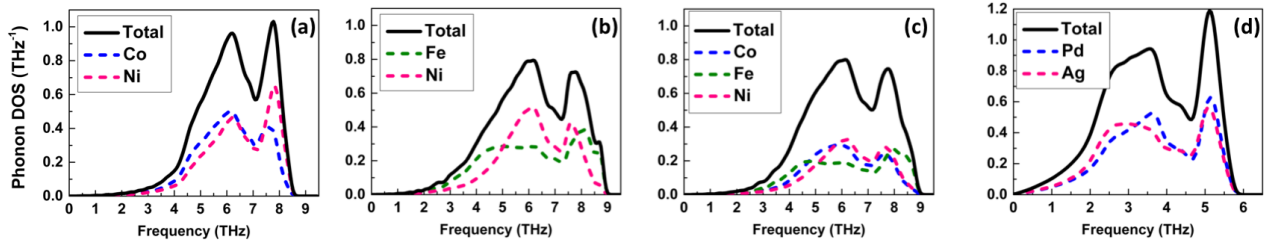

**Supplementary Figure 5** Total phonon DOS (solid black curves) and partial phonon DOS (colored dashed curves, see legends) in (a) NiCo, (b) NiFe, (c) NiFeCo and (d) AgPd from a single 108-atom SQS calculation.

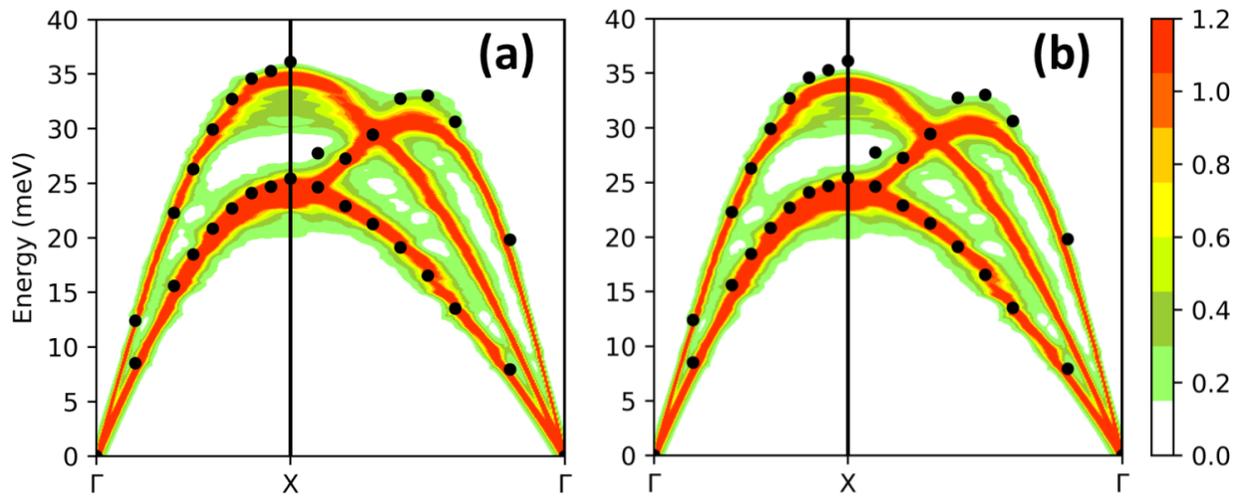

**Supplementary Figure 6** The phonon spectral function showing the dispersion and line widths of (a) the actual NiFe and (b) an "artificial NiCo" cell with the Fe mass replaced by the Co mass. Arbitrary units are used for the phonon spectral functions. The solid black circles represent the experimental measurements for NiFe.

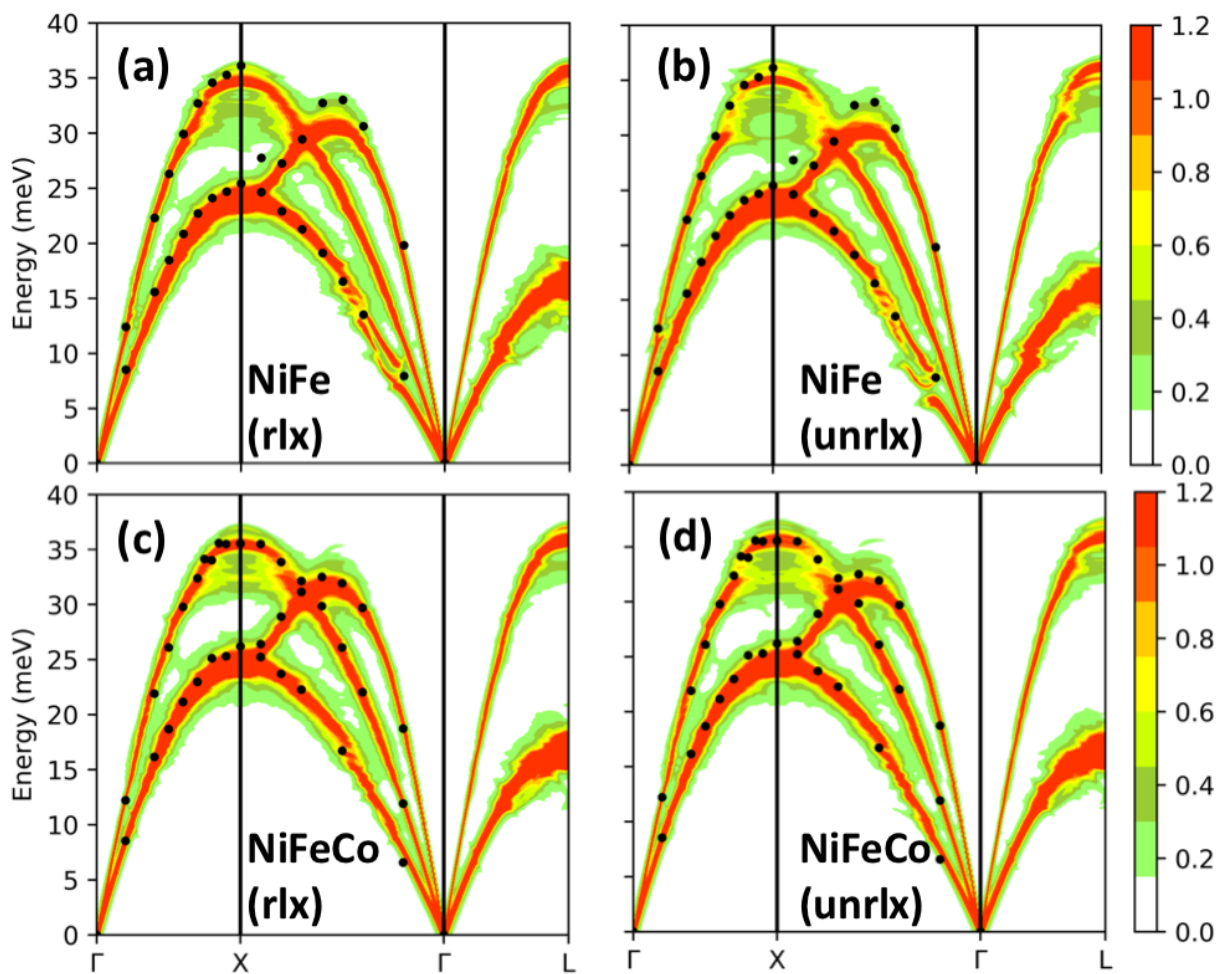

**Supplementary Figure 7** Phonon spectral function of NiFe and NiFeCo with local relaxation ((a), (c)) and without local relaxation ((b), (d)) demonstrating that relaxations have little impact.

# Tables

**Supplementary Table 1** The averaged force constants in NiCo, extracted from supercell calculations. The force constants are given in eV/Å$^2$. Values in parentheses are the standard deviations (eV/Å$^2$) of the corresponding species-pair force constant distribution.

|  |  | Ni-Ni | Ni-Co | Co-Co |
|---|---|---|---|---|
| 1st | α(xx = yy) | -0.917 (0.08) | -0.768 (0.11) | -0.635 (0.10) |
|  | β(xy = yx) | -0.995 (0.12) | -0.884 (0.13) | -0.796 (0.13) |
|  | γ(zz) | -0.081 (0.04) | 0.066 (0.05) | -0.005 (0.07) |
| 2nd | α(xx) | -0.167 | -0.218 | -0.264 |
|  | γ(yy = zz) | 0.021 | 0.038 | 0.069 |
| 3rd | α(xx) | -0.103 | -0.115 | -0.127 |
| | β(xy = yx) | -0.045 | -0.050 | -0.052 |
| | γ(yy = zz) | -0.043 | -0.043 | -0.036 |
| | δ(yz = zy) | 0.010 | 0.007 | -0.012 |
| 4th | α(xx = yy) | -0.106 | -0.145 | -0.209 |
| | β(xy = yx) | -0.080 | -0.121 | -0.184 |
| | γ(zz) | -0.011 | -0.013 | -0.013 |

**Supplementary Table 2** The averaged force constants in NiFe, extracted from supercell calculations. The force constants are given in eV/Å$^2$. Values in parentheses are the standard deviations (eV/Å$^2$) of the corresponding species-pair force constant distribution.

| | | Ni-Ni | Ni-Fe | Fe-Fe |
|---|---|---|---|---|
| 1st | α(xx = yy) | -0.771 (0.12) | -0.913 (0.12) | -1.187 (0.12) |
| | β(xy = yx) | -0.807 (0.10) | -1.031 (0.15) | -1.328 (0.17) |
| | γ(zz) | -0.054 (0.06) | -0.017 (0.07) | 0.099 (0.08) |
| 2nd | α(xx) | -0.162 | -0.070 | 0.112 |
| | γ(yy = zz) | 0.010 | 0.024 | 0.024 |
| 3rd | α(xx) | -0.090 | -0.070 | -0.045 |
| | β(xy = yx) | -0.037 | -0.031 | -0.023 |
| | γ(yy = zz) | -0.034 | -0.025 | -0.009 |
| | δ(yz = zy) | 0.002 | 0.002 | -0.002 |
| 4th | α(xx = yy) | -0.051 | -0.059 | -0.026 |
| | β(xy = yx) | -0.045 | -0.057 | -0.025 |
| | γ(zz) | -0.012 | 0.004 | 0.014 |

**Supplementary Table 3** The averaged force constants in AgPd, extracted from supercell calculations. The force constants are given in eV/Å$^2$. Values in parentheses are the standard deviations (eV/Å$^2$) of the corresponding species-pair force constant distribution.

| | | Ag-Ag | Ag-Pd | Pd-Pd |
|---|---|---|---|---|
| 1st | α(xx = yy) | -0.835 (0.10) | -0.873 (0.11) | -0.817 (0.10) |
| | β(xy = yx) | -0.967 (0.12) | -0.989 (0.13) | -0.926 (0.13) |
| | γ(zz) | 0.141 (0.02) | 0.100 (0.02) | 0.008 (0.03) |
| 2nd | α(xx) | -0.044 | -0.035 | -0.059 |
| | γ(yy = zz) | 0.022 | 0.029 | 0.036 |
| 3rd | α(xx) | -0.015 | -0.014 | -0.022 |
| | β(xy = yx) | -0.009 | -0.010 | -0.013 |
| | γ(yy = zz) | 0.002 | 0.000 | -0.009 |

|     |            | -0.002 | 0.002 | 0.007 |
|-----|------------|--------|-------|-------|
|     | δ(yz = zy) | -0.002 | 0.002 | 0.007 |
| 4th | α(xx = yy) | -0.007 | -0.011 | -0.001 |
|     | β(xy = yx) | -0.016 | -0.023 | -0.024 |
|     | γ(zz)      | 0.009  | 0.015 | 0.025 |

**Supplementary Table 4** The averaged force constants in NiFeCo, extracted from supercell calculations. The force constants are given in eV/Å². Values in parentheses are the standard deviations (eV/Å²) of the corresponding species-pair force constant distribution.

|     |            | Ni-Ni | Ni-Fe | Fe-Fe | Ni-Co | Fe-Co | Co-Co |
|-----|------------|-------|-------|-------|-------|-------|-------|
| 1st | α(xx = yy) | -0.814 (0.11) | -0.939 (0.13) | -1.235 (0.12) | -0.780 (0.13) | 0.903 (0.12) | -0.749 (0.17) |
|     | β(xy = yx) | -0.839 (0.11) | -1.071 (0.18) | -1.399 (0.18) | -0.850 (0.14) | -1.071 (0.17) | -0.894 (0.21) |
|     | γ(zz)      | -0.063 (0.05) | 0.008 (0.06) | 0.115 (0.06) | -0.022 (0.06) | 0.034 (0.07) | 0.001 (0.08) |
| 2nd | α(xx)      | -0.162 | -0.062 | 0.108 | -0.172 | 0.033 | -0.101 |
|     | γ(yy = zz) | -0.003 | 0.003 | 0.008 | 0.011 | 0.011 | 0.000 |
| 3rd | α(xx)      | -0.083 | -0.068 | -0.036 | -0.087 | -0.054 | -0.076 |
|     | β(xy = yx) | -0.035 | -0.028 | -0.016 | -0.035 | -0.020 | -0.028 |
|     | γ(yy = zz) | -0.034 | -0.029 | -0.010 | -0.035 | -0.022 | -0.031 |
|     | δ(yz = zy) | 0.005 | 0.003 | 0.000 | 0.001 | 0.004 | -0.002 |
| 4th | α(xx = yy) | -0.072 | -0.074 | -0.059 | -0.084 | -0.063 | -0.113 |
|     | β(xy = yx) | -0.061 | -0.070 | -0.056 | -0.075 | -0.061 | -0.108 |
|     | γ(zz)      | -0.015 | 0.004 | 0.015 | -0.003 | 0.011 | 0.006 |

**Supplementary Table 5** The averaged ($\langle \Phi_\parallel^{1st} \rangle$) and standard deviation ($\sigma \langle \Phi_\parallel^{1st} \rangle$) of the species-resolved force constants, given in units of eV/Å², in NiCo, NiFe, NiFeCo, AgPd (with and without local relaxation).

|   | NiCo | | | NiFe | | | AgPd | | |
|---|------|---|---|------|---|---|------|---|---|
|   | Ni-Ni | Ni-Co | Co-Co | Ni-Ni | Ni-Fe | Fe-Fe | Ag-Ag | Ag-Pd | Pd-Pd |
| $\langle \Phi_\parallel^{1st} \rangle$ | 0.956 | 0.825 | 0.716 | 0.789 | 0.972 | 1.258 | 0.902 | 0.931 | 0.870 |

| $\sigma\langle\Phi_\parallel^{1st}\rangle$ | 0.079 | 0.107 | 0.116 | 0.105 | 0.134 | 0.137 | 0.108 | 0.118 | 0.116 |
|---|---|---|---|---|---|---|---|---|---|
| | \multicolumn{6}{c|}{NiFeCo} | \multicolumn{3}{c|}{AgPd (no relax)} | | | | |
| | Ni-Ni | Ni-Fe | Fe-Fe | Ni-Co | Co-Fe | Co-Co | Ag-Ag | Ag-Pd | Pd-Pd |
| $\langle\Phi_\parallel^{1st}\rangle$ | 0.826 | 1.004 | 1.317 | 0.817 | 0.989 | 0.818 | 0.952 | 0.904 | 0738 |
| $\sigma\langle\Phi_\parallel^{1st}\rangle$ | 0.104 | 0.149 | 0.141 | 0.131 | 0.144 | 0.187 | 0.006 | 0.010 | 0.029 |